# Using Clustering Method to Understand Indian Stock Market Volatility

Tamal Datta Chaudhuri
Principal, Calcutta Business School, Diamond Harbour Road, Bishnupur – 743503, 24 Paraganas (South), West Bengal

Indranil Ghosh
Assistant Professor, Calcutta Business School, Diamond Harbour Road, Bishnupur – 743503, 24 Paraganas (South), West Bengal

## ABSTRACT
In this paper we use "Clustering Method" to understand whether stock market volatility can be predicted at all, and if so, when it can be predicted. The exercise has been performed for the Indian stock market on daily data for two years. For our analysis we map number of clusters against number of variables. We then test for efficiency of clustering. Our contention is that, given a fixed number of variables, one of them being historic volatility of NIFTY returns, if increase in the number of clusters improves clustering efficiency, then volatility cannot be predicted. Volatility then becomes random as, for a given time period, it gets classified in various clusters. On the other hand, if efficiency falls with increase in the number of clusters, then volatility can be predicted as there is some homogeneity in the data. If we fix the number of clusters and then increase the number of variables, this should have some impact on clustering efficiency. Indeed if we can hit upon, in a sense, an optimum number of variables, then if the number of clusters is reasonably small, we can use these variables to predict volatility. The variables that we consider for our study are volatility of NIFTY returns, volatility of gold returns, India VIX, CBOE VIX, volatility of crude oil returns, volatility of DJIA returns, volatility of DAX returns, volatility of Hang Seng returns and volatility of Nikkei returns. We use three clustering algorithms namely Kernel K-Means, Self-Organizing Maps and Mixture of Gaussian models and two internal clustering validity measures, Silhouette Index and Dunn Index, to assess the quality of generated clusters.

## 1. INTRODUCTION
The reasons for studying stock market volatility are that it i) aids in intraday trading, ii) is the basis of neutral trading in the options market, iii) affects portfolio rebalancing by fund managers, iv) helps in hedging, v) affects capital budgeting decisions through timing of raising equity from the market and its pricing and also vi) affects policy decisions relating to the financial markets. Extensive research has been done on stock market volatility and its implications, the thrust being on forecasting volatility. The measures that have been used for estimating volatility are historic volatility and implied volatility.

The literature has used econometric techniques like ARCH, GARCH models to estimate volatility. Using the mean reversal property of volatility, researchers have used decile analysis to predict volatility. This is useful for options traders. There has been application of Artificial Neural Network (ANN) models to forecast stock market volatility. This paper explores the role of Clustering Algorithms in forecasting volatility. We go beyond simply forecasting volatility and ask the question as to whether stock market volatility can be predicted at all and if so, within what time bounds. That is, whether it is meaningful to take a long time series data and predict volatility, without understanding the pattern in the data and its characteristics.

If we focus on implied volatility and study the data on India VIX, the implied volatility index in India, for the period 2008 to 2015 (June), Figure 1 shows that there is no specific trend or pattern in this data for long term forecasting. There are spikes in the data, and if we club the entire data for our analysis, we may be erring. Instead we suggest Clustering Algorithms in this paper to identify patterns in the data. For our analysis we map number of clusters against number of variables. We then test for efficiency of clustering. Our contention is that, given a fixed number of variables, one of them being historic volatility of NIFTY returns, if increase in the number of clusters improves clustering efficiency, then volatility cannot be predicted. Volatility then becomes random as, for a given time period, it gets classified in various clusters. On the other hand, if efficiency falls with increase in the number of clusters, then volatility can be predicted as there is some homogeneity in the data. Further, if we fix the number of clusters and then increase the number of variables, this should have some impact on clustering efficiency. Indeed if we can hit upon, in a sense, an optimum number of variables, then if the number of clusters is reasonably small, we can use these variables to predict volatility.

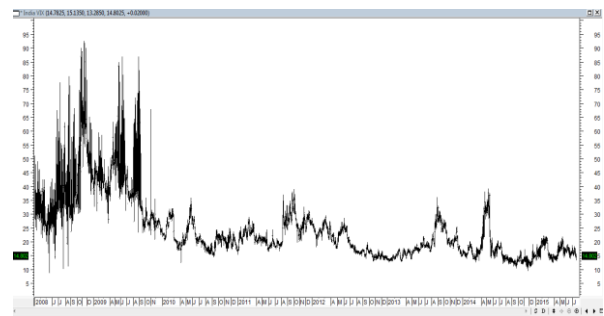

**Figure 1: India VIX for the period of 2008-2015 (June)
Source: Metastock**

## 2. OBJECTIVE OF THE STUDY

The objective of this paper is to present a framework of analysis based on Clustering Algorithms to forecast stock market volatility. The variables that we consider for our study are volatility of NIFTY returns, volatility of gold returns, India VIX, CBOE VIX, volatility of crude oil returns, volatility of DJIA returns, volatility of DAX returns, volatility of Hang Seng returns and volatility of Nikkei returns. Three clustering algorithms namely Kernel K-Means, Self-Organizing Maps and Mixture of Gaussian models will be used to carry out the clustering operation and two internal clustering validity measures, Silhouette Index and Dunn Index, will be computed to assess the quality of generated clusters. Although the purpose is to predict stock market volatility in India given by historic volatility of NIFTY returns with the help of the predictors mentioned above, our study is an exploration of patterns in the data to understand whether volatility can be predicted at all.

Accordingly, the plan of the paper is as follows. Section 3 explains the methodology of the study and a literature review is presented in Section 4. The variables are explained in Section 5 and Section 6 presents the results. Some concluding observations are made in Section 7.

## 3. METHODLOGY

Clustering is the process of partitioning the data objects into segments of homogeneous data objects based on similarity of some features. Each segment is known as a cluster. Objects belonging to a particular cluster are similar to one another and dissimilar to objects belonging to other clusters. It is an unsupervised learning process as no prior information about the class of data objects is available. Meaningful knowledge can only be inferred once the given set of data points is grouped into different clusters. Mathematically in N-dimensional Euclidean space, the task of clustering is to partition a given set (S) of data points $\{x_1, x_2, x_3, \ldots, x_n\}$ into K clusters $\{C_1, C_2, C_3, \ldots, C_n\}$ where the following conditions are satisfied:

$C_i \neq \emptyset$ for i=1,2,...., K ................................(1)

$C_i \cap C_j = \emptyset$ for i=1,2,..., K; j=1,2,..., K and i $\neq$ j ....(2)

and $\bigcup_{i=1}^{K} C_i = S$ ................................................(3)

Different clustering algorithms such as partitioning, divisive, density based and spectral clustering have been proposed and discussed throughout the literature. Similarly based on the nature of assignment of an object to a particular cluster, clustering techniques are classified as soft, fuzzy and probabilistic clustering. Some algorithms require number of clusters to be defined beforehand, while some others adjust the number of clusters based on some statistical measures respectively. To analyze the outcome of clustering or to assess the quality of formed clusters, broadly three different measures, internal, external and relative measures for clustering validations are usually applied. External measures are supervised techniques that compare the outcome against some prior ground truth information or expert-specified knowhow. Whereas internal measures are completely unsupervised techniques which measure the goodness of results by determining how well the clusters are separated and how compact they are. The approach of relative measures is to compare different clusters obtained by different parameter setting of same algorithms. Brief descriptions of working principles of these algorithms are provided below.

### 3.1 Kernel K-Means

It is a generalization of popular K-Means algorithm that overcomes the bottlenecks of the latter one. K-Means, a simple yet effective clustering tool, suffers if the data objects are not linearly separable. K-Means algorithm also fails to detect clusters which are not convex shaped. To overcome this obstacle Kernel K-Means algorithm projects data points of input space to a high dimensional feature space by applying nonlinear transformation functions (Kernel functions). Subsequently it follows the same principle of K-Means clustering algorithm in feature space to detect clusters. This algorithm initially generates a kernel matrix ($K_{ij}$) using equation

$K(x_i, x_j) = \varphi(x_i)^T \varphi(x_j)$ ................ (4)

where $x_i$, $x_j$ are data points to be clustered in input space. Usually a kernel function $K(x_i, x_j)$ is used to carry out the inner products in the feature space without explicitly defining transformation $\varphi$. Table 1 displays few well studied kernel functions as reported in literature.

**Table 1: Kernel Functions**

| Radial Basis Kernel | $K(x_i, x_j) = \exp\left(-\|x_i - x_j\|^2 / 2\sigma^2\right)$ |
|---|---|
| Polynomial Kernel | $K(x_i, x_j) = \left(x_i^T x_j + \gamma\right)^\delta$ |
| Sigmoid Kernel | $K(x_i, x_j) = \tanh\left(\gamma\left(x_i^T x_j\right) + \theta\right)$ |

**Source: Authors' own construction**

The outline of Kernel K-Means algorithm is illustrated below.

Step 1: Compute the Kernel matrix and initialize K cluster ($C_1, C_2, \ldots, C_k$) Centers arbitrarily.

Step 2: For each point $x_n$ and every cluster $C_i$ compute

$\| \varphi(x_n) - m_i \|^2$

Step 3: Find $c^*(x_n) = \text{argmin} \left(\|\varphi(x_n) - m_i\|^2\right)$

Step 4: Update clusters as $C_i = \{x_n | c^*(x_n) = i\}$

Step 5: Repeat steps 1 - 4 until convergence.

### 3.2 Gaussian Mixture Model

It is a probabilistic clustering tool where the objective is to infer a set of probabilistic clusters which is most likely to generate the data set aimed to be clustered. If S be a set of m probabilistic clusters $(s_1, s_2, \ldots, s_m)$ with probability density function $(f_1, f_2, \ldots, f_m)$ and probabilities $w_1, w_2, \ldots, w_m$ respectively, then for any data point d, the probability that d is generated by cluster $s_i$ is given by $P(d|s_i) = w_i f_i(d)$. The probability that d is generated by the set S of clusters is computed as

$P(d|S) = \sum_{i=1}^{m} w_i f_i(d)$ ................ (5)

If the data points are generated independently for data set, $D = (d_1, d_2, \ldots, d_n)$, then

$P(D|S) = \prod_{j=1}^{n} P(d_j|S)$

$= \prod_{j=1}^{n} \sum_{i=1}^{m} w_i f_i(d_i)$ ……………(6)

The objective of probabilistic model based clustering is to find a set of S probabilistic clusters such that P(D|S) is maximized. If the probability distribution functions are assumed to be Gaussian then the approach is known as Gaussian Mixture model. A multivariate Gaussian distribution function is characterized by the mean vector and covariance matrix. These parameters are estimated by Expectation Maximization algorithm.

In general if the data objects and parameters of m distribution are denoted by D={$d_1,d_2,….,d_n$} and Θ={ $Θ_1, Θ_2,….., Θ_m$} then equation 5 may be expressed as

$P(d_i|Θ) = \sum_{i=1}^{m} w_i P_i(di|Θ_i)$ ………………………… (7)

$P_i(d_i|Θ_i)$ is the probability that di is generated from jth distribution using parameter $Θ_i$. Equation 6 can be rewritten as

$P(D|Θ) = \prod_{j=1}^{n} \sum_{i=1}^{m} w_i P_i(d_j|Θ_i)$ …………………….(8)

For Gaussian Mixture Model, the objective is to estimate the parameters (mean vector and covariance matrix) that maximize equation 8.

Probability Distribution function of Gaussian distribution function is given by the following formula

$P(d_i|Θ) = \frac{1}{(2\pi)^{l/2}\sqrt{|\Sigma|}} exp\left(-\frac{(d_i-\mu)^T\Sigma^{-1}(d_i-\mu)}{2}\right)$ …, (9)

Where $\mu$ and $\Sigma$ are the mean and co-variance matrix of Gaussian and $l$ is the dimension of object $d_i$.

In Gaussian Mixture Model, the objective is to estimate the parameters (mean and covariance matrix) by Expectation Maximization (EM) algorithm that maximizes equation 10.

$Θ^* = \arg_Θ \max P(D|Θ)$ ……………………………..(10)

Generally logP(D|Θ) is maximized because of easier computations.

$\log P(D|Θ) = \log(\prod_{j=1}^{n} P(d_j|\theta)) = \sum_{j=1}^{n} \log\left(\sum_{i=1}^{m} w_i P_i(d_j|Θ_i)\right)$ ……..(11)

An auxiliary objective function, Q is considered instead directly maximizing the log likelihood.

$Q = \sum_{j=1}^{n} \sum_{i=1}^{m} \alpha_{ij} \log[w_i P_i(d_j|Θ_i)]$ ………(12)

Where $\alpha_{ij}$ is the respective posteriori probabilities for individual class i.

$\alpha_{ij} = \frac{w_i P_i(d_j|Θ_i)}{\sum_{r=1}^{k} P_r(d_j|Θ_r)}$ …………………(13)

$\sum_{i=1}^{n} \alpha_{ij} = 1$ ………………...(14)

Maximizing equation ensures P(D|Θ) is maximized if performed by an EM algorithm. The steps of EM algorithm is given below

E-Step: Compute 'expected' classes of all data points for each class using Equation 7.

M-Step: Maximum likelihood given the data's class membership distributions is computed by the following equations.

$W_i^{new} = \frac{1}{m} \sum_{j=1}^{m} \alpha_{ij}$ ………………(15)

$\mu_i^{new} = \frac{\sum_{j=1}^{m} \alpha_{ij} d_j}{\sum_{j=1}^{m} \alpha_{ij}}$ ………………..(16)

$\Sigma_i^{new} = \frac{\sum_{j=1}^{m} \alpha_{ij}(d_j-\mu_i^{new})(d_j-\mu_i^{new})^T}{\sum_{j=1}^{m} \alpha_{ij}}$ (17)

### 3.3 Self-Organizing Map

Self-Organizing Maps (SOM) belong to nonlinear Artificial Neural Network models proposed by Kohonen (1990). It is an unsupervised learning algorithm mainly deployed to reduce dimensions of data set and to find homogenous groupings (clusters) among the data points. It basically attempts to visualize high dimensional data patterns onto one or two dimensional grid or lattice of units (neurons) adaptively in a topological ordered manner. This transformation tries to preserve topological relations, i.e., patterns which are similar in the input space will be mapped to units that are close in the output space as well, and vice-versa. The units are connected to adjacent ones by a neighborhood relation which is varied dynamically in the network training process. The number of neurons accounts for the accuracy and generalization capability of the SOM. All neurons compete for each input pattern; the neuron that is chosen for the input pattern wins it. Only the winning neuron is activated (winner-takes-all). The winning neuron updates itself and neighbor neurons to approximate the distribution of the patterns in the input dataset. After the adaptation process is complete, similar clusters will be close to each other (i.e., topological ordering of clusters). The SOM network organizes itself by competing representation of the samples. Neurons are also allowed to change themselves in hoping to win the next competition. This selection and learning process makes the weights to organize themselves into a map representing similarities. The three key steps to form self-organizing maps are known as completion phase (identifying the best matching or winning neuron), cooperation phase (determining the location of topological neighborhood) and synaptic adaptation phase (self-organized formation of feature maps). The SOM algorithm is summarized below:

1. Initialization: Randomly initialize the weight vectors $W_j(0)$, where j = 1,2,……,l and l is the number of neurons in grid.

2. Sampling: Draw a sample training input vector x from the input space.

3. Similarity Matching: Find the best matching (winning) neuron i(x) at time step n by using minimum-distance criterion.

$$i(x) = \arg\min_j ||x(n) - w_j||, \ j = 1,2,………,l$$

4. Updating: Adjust the synaptic-weight vectors of all excited neurons

$$w_j(n+1) = w_j(n) + \varepsilon(n)\ h_{j,i(x)}(n)(x(n)-w_j(n)),$$

where $\varepsilon(n)$ is learning rate and $h_{j,i(x)}(n)$ is the neighborhood function centered around i(x), the winning or best matching unit. In this study neighborhood function is computed as

$$h_{j,i(x)}(n) = \exp\left(-\frac{d_{j,i}^2}{2\sigma^2}\right)$$

Parameter $\sigma$ is the effective width of the topological neighborhood.

5. Continuation: Repeat step 2-4 until convergence.

Due to unavailability of ground truth information, we have opted for internal clustering validity index measures. Basically they evaluate a clustering by analyzing separation of and compactness of individual clusters. These indices sometimes are also applied to automatically determine the number of clusters. However, in this study instead of fixing the number of clusters, these measures are computed across a range of number of clusters as the objective is to infer the nature of stock market volatility. Silhouette Index (SI), Dunn Index (DI), Alternative Dunn index (ADI), Krzanowski–Lai index (KL) and Calinski–Harabasz index (CH) are examples of various internal validation measures which have been used frequently in different applications reported in literature. Here we have employed Silhouette Index (SI) and Dunn Index (DI) separately to assess the clustering results.

### 3.4 Silhouette Index (SI)
For a dataset D of n objects, if D is partitioned into k clusters, $C_1,…,C_K$, Silhouette Index, s(i) for each object i∈D is computed as

$$s(i) = \frac{b(i)-a(i)}{max\{a(i),b(i)\}}$$

Here $a(i)$ is the average distance between i and all other objects in the cluster in which i belongs whereas $b(i)$ is the minimum average distance from i to all clusters to which i does not belong. The Value of SI ranges between -1 and 1. A larger value indicates better quality clustering result.

### 3.5 Dunn Index (DI)
DI is a ratio between the minimal inter cluster distance to maximal intra cluster distance. The index is computed as

$$D = \frac{d_{min}}{d_{max}}$$

where $d_{min}$ represents the smallest distance between two objects from different clusters and $d_{max}$ denotes the largest distance of two objects from the same cluster. Larger value of DI implies better quality clusters.

## 4. LITERATURE REVIEW
Clustering is an active area of data mining research and many applications in the area of image and video processing, telecommunication churn management, stock market analysis, system biology, social network analysis and cellular manufacturing have been reported in the literature. Ozer (2001) utilized fuzzy clustering analysis for user segmentation of online music services. Nanda et al. (2010) adopted K-Means, Self-Organizing Maps and Fuzzy C-Means based clustering algorithm to classify Indian stocks in different clusters and subsequently developed portfolios from these clusters. Kim and Ahn (2008) applied a Genetic Algorithm based K-Means clustering algorithm to develop recommender system for online shopping market. Siyal and Yu (2005) proposed a modified FCM algorithm for bias (also called intensity in-homogeneities) estimation and segmentation of MR (Magnetic resonance) images. Sun and Wing (2005) utilized K-Means algorithm to study the effect and implementation of different critical success factors for new product development in Hong Kong toy industry. Chattopadhyay et al. (2011) proposed a novel framework based on principal component analysis (PCA) and Self-Organizing Map (SOM) to carry out automatic cell formation in cellular manufacturing layouts.

Apart from applications based studies, significant amount of research work has been dedicated towards fundamental development of clustering methods. Maulik and Bandyopadhyay (2000) introduced genetic algorithm based clustering algorithm which displayed performance superiority over K-Means algorithm on artificial and real life data sets. Mitra et al. (2010) proposed a new clustering technique, Shadowed C-Means, integrating fundamental principles of fuzzy and rough clustering techniques. Later Mitra et al. (2011) utilized this algorithm for satellite image segmentation. Ju and Liu (2010) introduced fuzzy Gaussian mixture model (FGMM) based clustering hybridizing conventional Gaussian Mixture Model and Fuzzy set theory for faster convergence and tackling nonlinear data set. Hatamlou (2012) developed a new heuristic optimization based clustering technique, Black Hole algorithm, which outperformed several standard clustering methods. Chaira et al. (2011) proposed a new Intuitionistic Fuzzy C-Means algorithm defined on intuitionistic fuzzy set and successfully applied it to cluster CT scan brain images. There are many other clustering algorithms namely, Neural Gas, Artificial Bee Colony Based Clustering technique (ABC), Gravitational search approach (GSA), Particle Swarm Optimization (PSO) based approach, Ant Colony Optimization (ACO) based technique, Chameleon and DBSCAN that have been reported in literature.

Application of Decile Analysis in forecasting stock market volatility can be seen in McMillan (2004) and Datta Chaudhuri and Sheth (2014). The literature on volatility prediction by the ARCH/GARCH method includes papers by Das and Bhattacharya (2014), Karolyi (1995), Kumar and Mukhopadhyay (2007), Angela (2000), and Padhi and Logesh (2012). Datta Chaudhuri and Ghosh (2015), deployed

Artificial Neural Network based framework for prediction of stock market volatility in the Indian stock market through volatility of NIFTY returns and volatility of gold returns.

## 5. THE VARIABLES

For our analysis we have considered daily data of nine variables namely volatility of NIFTY returns (NIFTYSDR), volatility of gold returns (GOLDSDR), India VIX, CBOE VIX, volatility of crude oil returns (CRUDESDR), volatility of DJIA returns (DJIASDR), volatility of DAX returns (DAXSDR), volatility of Hang Seng returns (HANGSDR) and volatility of Nikkei returns (NIKKEISDR) for the years 2013 and 2014. In the analysis there are no inputs or outputs. All the variables are considered together to identify clusters. However, the implicit reason for choosing the variables is that there does exist some association between them and hence do play a role in explaining historic volatility. Figures 2 and 3 provide examples of two such long term associations

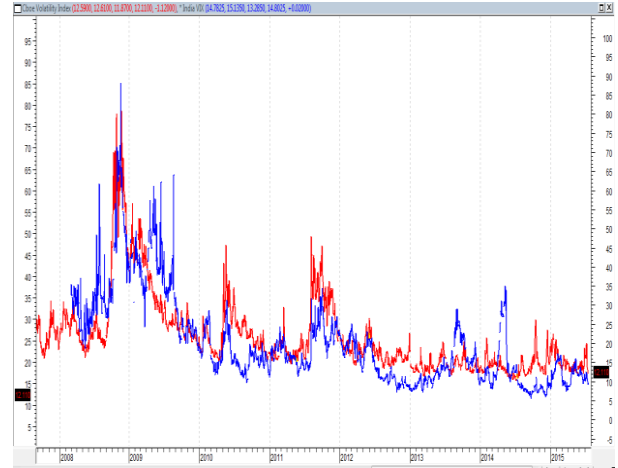

**Figure 3: INDIA VIX AND CBOE VIX FOR THE PERIOD 2008 – 2015 (June)**
**Source: Metastock**

Figure 2 indicates that, over a fairly long period, historic volatility and implied volatility do move together. So considering INDIA VIX as a predictor of NIFTYSDR is alright. Further, it may be observed from Figure 3 that expected volatility in the US seems to go hand in hand with expected volatility in India. That is, global uncertainties affect US implied volatility, which in turn affects implied volatility in India. To allow for external shocks, as India is a large importer of crude oil, we consider CRUDESDR in the analysis. In the recent past, political instability in the Middle East and related regions has impacted the expected availability of oil and has resulted in stock market instability in India. Global instability, both in the western and the eastern world has been incorporated through DJIASDR, DAXSDR, HANGSDR and NIKKEISDR.

## 6. RESULTS AND ANALYSIS

Tables 2 to 4 present the obtained DI values of the clusters generated by the three algorithms for different combinations of features and number of clusters.

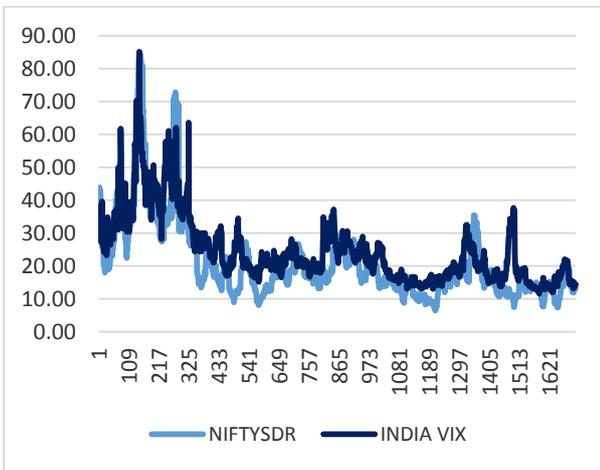

**Figure 2: INDIAVIX and NIFTYSDR for the period 3.3.2008 to 10.4.2015**
**Source: Authors' own construction**

**Table 2: DI values of clustering result generated by Kernel K-Means algorithm**

|  |  | No. of Features | | | | | | | |
|---|---|---|---|---|---|---|---|---|---|
|  |  | 2 | 3 | 4 | 5 | 6 | 7 | 8 | 9 |
| No. of Clusters | 2 | 0.0303 | 0.0573 | 0.0443 | 0.0447 | 0.0499 | 0.0926 | 0.0979 | 0.0976 |
|  | 3 | 0.0322 | 0.102 | 0.107 | 0.0509 | 0.0697 | 0.0499 | 0.0859 | 0.0519 |
|  | 4 | 0.0282 | 0.0987 | 0.0497 | 0.0988 | 0.0561 | 0.0977 | 0.1248 | 0.0608 |
|  | 5 | 0.0813 | 0.0947 | 0.0473 | 0.099 | 0.1363 | 0.1361 | 0.1454 | 0.1451 |
|  | 6 | 0.0841 | 0.0598 | 0.0476 | 0.0831 | 0.0913 | 0.0795 | 0.1251 | 0.1222 |
|  | 7 | 0.0327 | 0.0845 | 0.0954 | 0.0831 | 0.1632 | 0.1686 | 0.0985 | 0.0718 |
|  | 8 | 0.0841 | 0.0889 | 0.0425 | 0.0736 | 0.126 | 0.1419 | 0.1108 | 0.1102 |
|  | 9 | 0.0813 | 0.0928 | 0.0918 | 0.0898 | 0.126 | 0.1563 | 0.1108 | 0.1102 |
|  | 10 | 0.0867 | 0.1561 | 0.0993 | 0.0909 | 0.126 | 0.1285 | 0.1228 | 0.1102 |
|  | 11 | 0.0661 | 0.1059 | 0.0993 | 0.1687 | 0.133 | 0.1285 | 0.1234 | 0.1186 |

**Source: Authors' own construction**

**Table 3: DI values of clustering result generated by Self-Organizing Map**

|  |  | No. of Features | | | | | | | |
|---|---|---|---|---|---|---|---|---|---|
|  |  | 2 | 3 | 4 | 5 | 6 | 7 | 8 | 9 |
| No. of Clusters | 2 | 0.0237 | 0.0515 | 0.0298 | 0.0198 | 0.0571 | 0.0499 | 0.0979 | 0.0976 |
|  | 3 | 0.0339 | 0.102 | 0.039 | 0.1309 | 0.0392 | 0.0344 | 0.0764 | 0.0875 |
|  | 4 | 0.0282 | 0.0987 | 0.1039 | 0.0596 | 0.0919 | 0.0557 | 0.0723 | 0.0671 |
|  | 5 | 0.0476 | 0.0348 | 0.0473 | 0.099 | 0.0754 | 0.0717 | 0.0817 | 0.0987 |
|  | 6 | 0.0421 | 0.0743 | 0.0482 | 0.0831 | 0.1247 | 0.1279 | 0.0913 | 0.081 |
|  | 7 | 0.0501 | 0.0606 | 0.0461 | 0.0681 | 0.1632 | 0.1686 | 0.0985 | 0.1518 |
|  | 8 | 0.0379 | 0.0603 | 0.0435 | 0.0987 | 0.0959 | 0.1119 | 0.1108 | 0.1102 |
|  | 9 | 0.0545 | 0.0889 | 0.0918 | 0.086 | 0.126 | 0.1122 | 0.0669 | 0.109 |
|  | 10 | 0.0545 | 0.1427 | 0.096 | 0.0718 | 0.1423 | 0.1113 | 0.0908 | 0.0907 |
|  | 11 | 0.0578 | 0.1463 | 0.0994 | 0.0909 | 0.0723 | 0.1807 | 0.0829 | 0.0907 |

**Source: Authors' own construction**

**Table 4: DI values of clustering result generated by Gaussian Mixture Model**

|  |  | No. of Features | | | | | | | |
|---|---|---|---|---|---|---|---|---|---|
|  |  | 2 | 3 | 4 | 5 | 6 | 7 | 8 | 9 |
| No. of Clusters | 2 | 0.0103 | 0.1456 | 0.0473 | 0.0317 | 0.0313 | 0.1436 | 0.0675 | 0.1816 |
|  | 3 | 0.0048 | 0.0749 | 0.0359 | 0.228 | 0.0587 | 0.1436 | 0.0744 | 0.0541 |
|  | 4 | 0.0272 | 0.0592 | 0.0208 | 0.2135 | 0.0188 | 0.0495 | 0.0744 | 0.1152 |
|  | 5 | 0.0379 | 0.0377 | 0.091 | 0.2796 | 0.0722 | 0.0729 | 0.0858 | 0.1152 |
|  | 6 | 0.048 | 0.0288 | 0.0526 | 0.3079 | 0.0837 | 0.0985 | 0.0858 | 0.1428 |
|  | 7 | 0.0377 | 0.0403 | 0.0441 | 0.3581 | 0.0837 | 0.0985 | 0.0925 | 0.0687 |
|  | 8 | 0.0449 | 0.054 | 0.0701 | 0.3482 | 0.1095 | 0.0985 | 0.0925 | 0.0687 |
|  | 9 | 0.0209 | 0.0362 | 0.0309 | 0.3183 | 0.0954 | 0.0985 | 0.1033 | 0.1307 |
|  | 10 | 0.0288 | 0.0478 | 0.0475 | 0.3533 | 0.0954 | 0.0985 | 0.1318 | 0.0903 |
|  | 11 | 0.0499 | 0.0387 | 0.1036 | 0.3808 | 0.112 | 0.1748 | 0.1593 | 0.111 |

**Source: Authors' own construction**

In Table 2, the maximum DI value of 0.1686 corresponds to 5 features (India VIX, NIFTYSDR, CBOE VIX, volatility of crude oil returns, volatility of DJIA returns) and 7 clusters. Similarly, in Tables 3 and 4, maximum DI values correspond to 7 features (India VIX, NIFTYSDR, CBOE VIX, volatility of crude oil returns, volatility of DJIA returns, volatility of DAX returns, volatility of Hang Seng returns) 11 clusters and 5 features (India VIX, NIFTYSDR, CBOE VIX, volatility of crude oil returns, volatility of DJIA returns) 11 clusters respectively. For better understanding, following figures map the relationship between number of features and number of clusters. Five common features present in all three experiments are India VIX, NIFTYSDR, CBOE VIX, volatility of crude oil returns and volatility of DJIA returns.

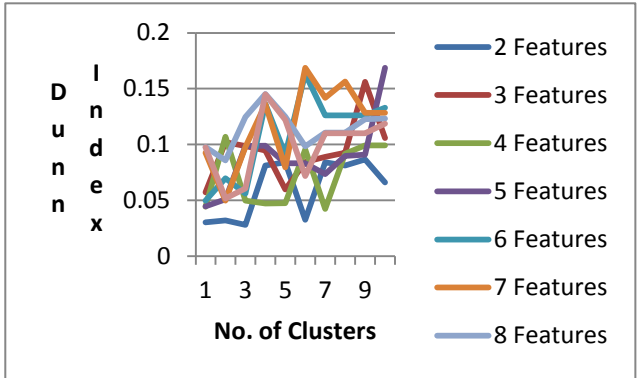

**Figure 4: DI values of clustering/features generated by Kernel K-Means algorithm**
**Source: Authors' own construction**

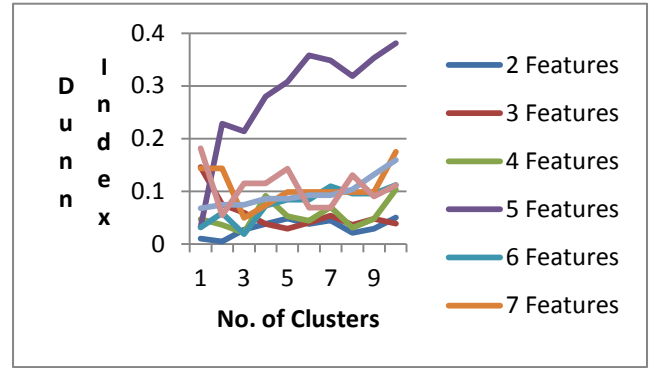

**Figure 6: DI values of clustering/features generated by Self Organizing Maps**
**Source: Authors' own construction**

As the Figure 1 contains several spikes (both in positive and negative direction corresponding to local maxima and minima) it is hard to determine whether incremental increase in number clusters result in good or bad quality clusters. However, it may be broadly inferred that large number of features (6-9) produces better quality segmentation in compared to smaller number of features (1-3). Figure 2 justifies the claim as well. Figure 3 clearly identifies that usage of 5 features (India VIX, NIFTYSDR, CBOE VIX, volatility of crude oil returns, volatility of DJIA returns) yields superior cluster quality than other combinations.

Same clustering algorithms are applied on the same data set to calculate Silhouette Index values. Results are summarized in tables 5-7.

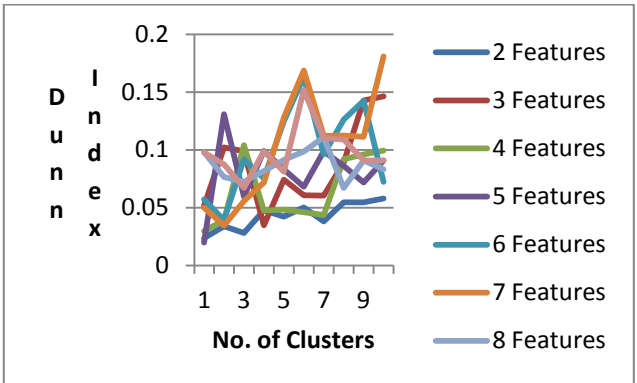

**Figure 5: DI values of clustering/features generated by Gaussian Mixture Model**
**Source: Authors' own construction**

**Table 5: SI values of clustering result generated by Kernel K-Means algorithm**

|  |  | No. of Features | | | | | | | |
|---|---|---|---|---|---|---|---|---|---|
|  |  | 2 | 3 | 4 | 5 | 6 | 7 | 8 | 9 |
| No. of Clusters | 2 | 0.5914 | 0.4809 | 0.4062 | 0.3786 | 0.356 | 0.3446 | 0.3447 | 0.3372 |
|  | 3 | 0.5994 | 0.5147 | 0.4525 | 0.4278 | 0.3912 | 0.3805 | 0.3331 | 0.3689 |
|  | 4 | 0.5527 | 0.4941 | 0.4183 | 0.4592 | 0.4268 | 0.4071 | 0.3811 | 0.3579 |
|  | 5 | 0.5659 | 0.4823 | 0.4601 | 0.4568 | 0.4606 | 0.4419 | 0.4168 | 0.4091 |
|  | 6 | 0.5121 | 0.4734 | 0.4461 | 0.4614 | 0.4868 | 0.4617 | 0.437 | 0.4284 |
|  | 7 | 0.4836 | 0.4156 | 0.4631 | 0.466 | 0.4731 | 0.4551 | 0.4395 | 0.4173 |
|  | 8 | 0.5266 | 0.4736 | 0.4494 | 0.4323 | 0.4669 | 0.4464 | 0.4446 | 0.4376 |

|  |  |  |  |  |  |  |  |  |
|---|---|---|---|---|---|---|---|---|
|  | 9 | 0.5129 | 0.4774 | 0.4575 | 0.4248 | 0.4169 | 0.4476 | 0.45 | 0.4232 |
|  | 10 | 0.4965 | 0.4681 | 0.4349 | 0.4157 | 0.4154 | 0.4481 | 0.4515 | 0.4351 |
|  | 11 | 0.4966 | 0.4378 | 0.4285 | 0.4207 | 0.4076 | 0.4383 | 0.4469 | 0.4326 |

**Source: Authors' own construction**

**Table 6: SI values of clustering result generated by Self-Organizing Map**

|  |  | No. of Features ||||||||
|---|---|---|---|---|---|---|---|---|---|
|  |  | 2 | 3 | 4 | 5 | 6 | 7 | 8 | 9 |
| No. of Clusters | 2 | 0.4733 | 0.4741 | 0.4043 | 0.4043 | 0.3568 | 0.345 | 0.3462 | 0.3372 |
|  | 3 | 0.5932 | 0.5147 | 0.4474 | 0.4257 | 0.3369 | 0.345 | 0.371 | 0.3691 |
|  | 4 | 0.5527 | 0.4941 | 0.4606 | 0.4577 | 0.4233 | 0.3791 | 0.3973 | 0.3944 |
|  | 5 | 0.5246 | 0.45 | 0.4601 | 0.4569 | 0.4458 | 0.4109 | 0.397 | 0.4129 |
|  | 6 | 0.4868 | 0.466 | 0.4155 | 0.4614 | 0.4297 | 0.4234 | 0.3883 | 0.372 |
|  | 7 | 0.4616 | 0.4582 | 0.4431 | 0.4309 | 0.472 | 0.4134 | 0.4381 | 0.4309 |
|  | 8 | 0.4525 | 0.4274 | 0.4431 | 0.4004 | 0.4643 | 0.4551 | 0.4446 | 0.4376 |
|  | 9 | 0.4875 | 0.4401 | 0.4575 | 0.4368 | 0.4637 | 0.4505 | 0.4299 | 0.4201 |
|  | 10 | 0.4778 | 0.4279 | 0.432 | 0.418 | 0.4382 | 0.3641 | 0.4366 | 0.409 |
|  | 11 | 0.4562 | 0.4221 | 0.4733 | 0.4111 | 0.4334 | 0.4411 | 0.4019 | 0.406 |

**Source: Authors' own construction**

**Table 7: SI values of clustering result generated by Gaussians Mixture Model**

|  |  | No. of Features ||||||||
|---|---|---|---|---|---|---|---|---|---|
|  |  | 2 | 3 | 4 | 5 | 6 | 7 | 8 | 9 |
| No. of Clusters | 2 | 0.4733 | 0.4741 | 0.4043 | 0.4043 | 0.3568 | 0.345 | 0.3462 | 0.3372 |
|  | 3 | 0.5932 | 0.5147 | 0.4474 | 0.4257 | 0.3369 | 0.345 | 0.371 | 0.3691 |
|  | 4 | 0.5527 | 0.4941 | 0.4606 | 0.4577 | 0.4233 | 0.3791 | 0.3973 | 0.3944 |
|  | 5 | 0.5246 | 0.45 | 0.4601 | 0.4569 | 0.4458 | 0.4109 | 0.397 | 0.4129 |
|  | 6 | 0.4868 | 0.466 | 0.4155 | 0.4614 | 0.4297 | 0.4234 | 0.3883 | 0.372 |
|  | 7 | 0.4616 | 0.4582 | 0.4431 | 0.4309 | 0.472 | 0.4134 | 0.4381 | 0.4309 |
|  | 8 | 0.4525 | 0.4274 | 0.4431 | 0.4004 | 0.4643 | 0.4551 | 0.4446 | 0.4376 |
|  | 9 | 0.4875 | 0.4401 | 0.4575 | 0.4368 | 0.4637 | 0.4505 | 0.4299 | 0.4201 |
|  | 10 | 0.4778 | 0.4279 | 0.432 | 0.418 | 0.4382 | 0.3641 | 0.4366 | 0.409 |
|  | 11 | 0.4562 | 0.4221 | 0.4733 | 0.4111 | 0.4334 | 0.4411 | 0.4019 | 0.406 |

**Source: Authors' own construction**

Maximum SI values of table 5, 6 and 7 correspond to 2 features (India VIX and NIFTYSDR) and 3 clusters (0.5994), 2 features (India VIX and NIFTYSDR) and 3 clusters (0.5932), 2 features (India VIX and NIFTYSDR) and 3 clusters (0.5932) respectively. Unlike the pattern observed in DI values, here it is quite evident that increase in number of clusters does not improve the quality of clusters. It also indicates that addition of extra features fails to enhance clusters quality significantly as well. The results are depicted in following figures.

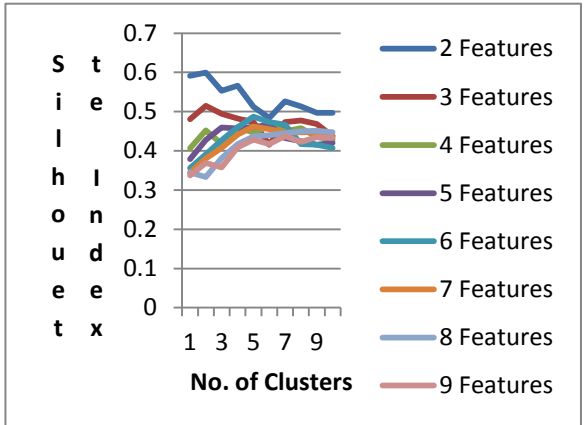

**Figure 7: SI Index obtained by Kernel K-Means technique**

Source: Authors' own construction

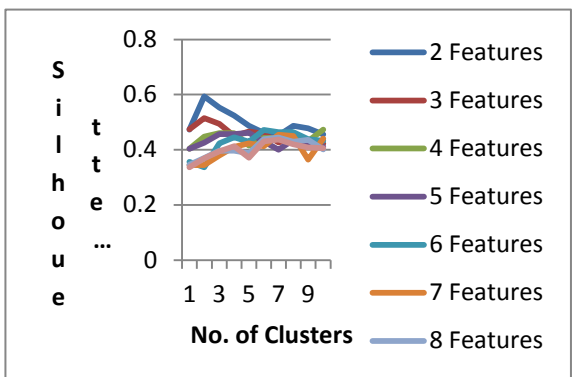

**Figure 8: SI Index obtained by Self-Organizing Map technique**

Source: Authors' own construction

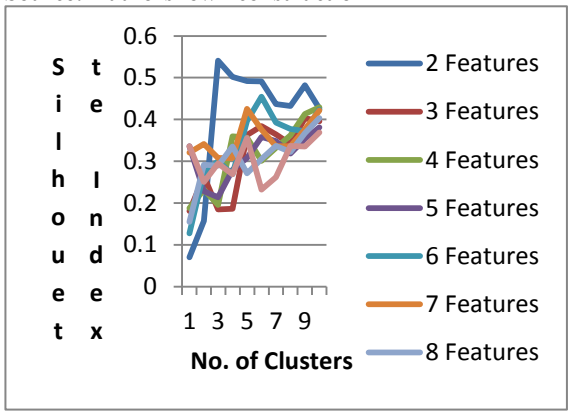

**Figure 9: SI Index obtained by Gaussian Mixture Model**

Source: Authors' own construction

## 7. CONCLUDING REMARKS

The purpose of this paper was to demonstrate that volatility prediction in stock markets has to be preceded by a study of the number of predictors and the number of clusters. The data on historic volatility may not be homogenous and the presence of many clusters would validate that. If there are too many clusters then it implies that volatility is random and would be difficult to predict. Further, the choice of the predictors has to be mapped with the number of clusters. Too many predictors with large number of clusters over a long time series data may not yield efficient results. Our study for two years for the Indian stock market reveals that of the variables chosen, seven predictors over five to six clusters gave optimum results. This implies that, given the time span as defined by a cluster, one can predict volatility with the help of the predictors. For data spanning across clusters, prediction may not be desirable. Diagrams of the Silhouette Index for the algorithms indicate that the data in the sample can at most be broken into three clusters. This implies that three broad distinct associations were seen among the variables chosen, and within the clusters forecasting is possible.

## 8. REFERENCES

[1] Angela, N., (2000), Volatility Spillover Effects from Japan and the US to Pacific-Basin, Journal of International Money and Finance, 19, 207-233.

[2] Chaira, T., (2011), A novel intuitionistic fuzzy C means clustering algorithm and its application to medical images, Applied Soft Computing, 11, 1711-1717.

[3] Chattopadhyay, M., Dan, P., K. & Mazumdar, S., (2011), Principal component analysis and self-organizing map for visual clustering of machine-part cell formation in cellular manufacturing system, Systems Research Forum, 5, 25-51.

[4] Das, S. & Bhattacharya, B., (2014), Global Financial Crisis and Pattern of Return and Volatility Spill-over from the Stock Markets of USA and Japan on the Indian Stock Market: An Application of EGARCH Model, CBS Journal of Management Practices, 1, 1-18.

[5] Datta Chaudhuri, T. and Kinjal, S., (2014), Forecasting Volatility, Volatility Trading and Decomposition by Greeks, CBS Journal of Management Practices, 1, 59-70.

[6] Datta Chaudhuri, T. & Ghosh, I. (2015), Forecasting Volatility in Indian Stock Market Using Artificial Neural Network with Multiple Inputs and outputs, International Journal of Computer Applications, 120, 7-15.

[7] Hatamlou, A., (2012), Black hole: A new heuristic optimization approach for data clustering, Information Sciences, 222, 175-184.

[8] Ju, Z. & Liu, H., (2012), Fuzzy Gaussian Mixture Models, Pattern Recognition, 45, 1146–1158.

[9] Karloyi, G., A., (1995), A Multivariate GARCH Model of International Transmissions of Stock Returns and Volatility: The Case of United States and Canada, Journal of Business and Economic Statistics, 31, 11-25.

[10] Kim, K.-j., & Ahn, H. (2008). A recommender system using GA K-means clustering in an online shopping market.Expert Systems with Applications, 34, 1200–1209.

[11] Kumar, K., K. & Mukhopadhyay, C. (2002), Volatility Spillovers from US to Indian Stock Market: A Comparison of GARCH Models, ICFAI Journal of Financial Economics, 5, 7-30.

[12] Maulik, U., & Bandyopadhyay, S., (2000), Genetic algorithm-based clustering technique, Pattern Recognition, 33, 1455-1465.

[13] McMillan, Lawrence G (2004), McMillan on Options, John Wiley & Sons, Inc., Hoboken, New Jersey.

[14] Mitra, S., Pedrycz, W. & Barman, B. (2010), Shadowed

c-means: Integrating fuzzy and rough clustering, Pattern Recognition, 43, 1282–1291.

[15] Mitra, S. & Kundu, P., P., (2011), Satellite image segmentation with Shadowed C-Means, Information Sciences, 181, 3601-3613.

[16] Nanda, S., R., Mahanty, B. & Tiwari, M., K., (2010), Clustering Indian stock market data for portfolio management, Expert Systems with Applications, 37, 8793–8798.

[17] Ozer, M., (2001), User segmentation of online music services using fuzzy clustering, Omega, 29, 193-206.

[18] Padhi, P. and Lagesh, M., A., (2012), Volatility Spillover and Time Varying Correlation Among the Indian and US Stock Markets, Journal of Quantitative Economics, 10.

[19] Sinha, P. and Sinha, G., (2010), Volatility Spillover in India, USA and Japan Investigation of Recession Effects, http://mpra.ub.uni-muenchen.de/47190/MPRA, paper No. 47190.

[20] Siyal, M., Y. & Yu, L., (2005), An intelligent modified fuzzy c-means based algorithm for bias estimation and segmentation of brain MRI, Pattern Recognition Letter, 26, 2052–2062.

[21] Sun, H. and Wing, W. C., (2005) Critical success factors for new product development in the Hong Kong toy industry, Technovation, 25, 293–303.